\documentclass[reprint, superscriptaddress, nofootinbib, pra]{revtex4} 
\usepackage{amssymb, amsmath}
\usepackage{graphicx}
\usepackage{xcolor}
\usepackage{csquotes}
\usepackage[outdir=./]{epstopdf}
\usepackage{verbatim}
\begin{document}
\title{Superdeterministic hidden-variables models I: non-equilibrium and signalling}
\author{Indrajit Sen} \email{isen@clemson.edu}
\affiliation{Department of Physics and Astronomy,\\
Clemson University, Kinard Laboratory,\\
Clemson, SC 29634, USA}
\author{Antony Valentini} \email{antonyv@clemson.edu}
\affiliation{Department of Physics and Astronomy,\\
Clemson University, Kinard Laboratory,\\
Clemson, SC 29634, USA}
\affiliation{Augustus College,\\
14 Augustus Road,\\
London SW19 6LN UK}
\date{\today}

\begin{abstract}
This is the first of two papers which attempt to comprehensively analyse superdeterministic hidden-variables models of Bell correlations. We first give an overview of superdeterminism and discuss various criticisms of it raised in the literature. We argue that the most common criticism, the violation of `free-will', is incorrect. We take up Bell's intuitive criticism that these models are `conspiratorial'. To develop this further, we introduce nonequilibrium extensions of superdeterministic models. We show that the measurement statistics of these extended models depend on the physical system used to determine the measurement settings. This suggests a fine-tuning in order to eliminate this dependence from experimental observation. We also study the signalling properties of these extended models. We show that although they generally violate the formal no-signalling constraints, this violation cannot be equated to an actual signal. We therefore suggest that the so-called no-signalling constraints be more appropriately named the marginal-independence constraints. We discuss the mechanism by which marginal-independence is violated in superdeterministic models. Lastly, we consider a hypothetical scenario where two experimenters use the apparent-signalling of a superdeterministic model to communicate with each other. This scenario suggests another conspiratorial feature peculiar to superdeterminism. These suggestions are quantitatively developed in the second paper.
\end{abstract}
\maketitle

\section{Introduction}
Orthodox quantum mechanics abandons realism in the microscopic world for an operationalist account consisting of macroscopic preparations and measurements performed by experimenters. This abandonment has led to several difficulties about the interpretation of quantum mechanics \cite{bell, solventinich}. Contrary to the expectations of many of the early practitioners of the theory, these difficulties have only grown more acute with time, as quantum mechanics has come to be applied to newer fields like cosmology and quantum gravity. One option to resolve these long-standing difficulties is to restore realism in the microscopic world. This restoration is the goal of hidden-variables reformulations of quantum mechanics. The general form of such a reformulation can be illustrated quite simply. Consider a quantum experiment where the system is prepared in a quantum state $\psi$ and a measurement $M$ (defined by a Hermitian operator) is subsequently performed upon it. Orthodox quantum mechanics predicts, for an ensemble, an outcome probability $p(k|\psi, M)$ for obtaining the $k^{th}$ outcome. This outcome probability can be expanded, using the standard rules of probability theory, as
\begin{align}
p(k|\psi, M) = \int d\lambda p(k|\psi, M, \lambda) \rho(\lambda|\psi, M) \label{ola}
\end{align}
in terms of $\lambda$ which label the hidden-variables currently inaccessible to the experimenters. A hidden-variables model of this experiment must define, first, the hidden-variables $\lambda$, second, the distribution $\rho(\lambda|\psi, M)$ of $\lambda$'s over the ensemble, and third, the distribution $p(k|\psi, M, \lambda)$ of outcomes given a particular $\lambda$. Note that, in general, equation (\ref{ola}) involves a correlation between $\lambda$ and the future measurement setting $M$.\\

On the other hand, one of the most important results in the interpretation of quantum mechanics, Bell's theorem \cite{bell}, assumes there to be no correlation between the hidden variables and the measurement settings. Without this assumption, called `measurement-independence' \cite{hall10, howmuch, hall16, hall19} in recent literature, local hidden-variables models of quantum mechanics cannot be ruled out via Bell's theorem. The subject of this paper, and a subsequent one denoted by $\mathcal{B}$ \cite{2nd}, is superdeterminism. Superdeterministic models (for examples, see refs. \cite{brans, hautomaton, palmerend}) circumvent Bell's theorem by violating the measurement-independence assumption.\\

We define a superdeterministic model as one having the following two properties:\\
1. Determinism: Every event in the universe is determined given past conditions. In the present context, the specific implication is that the choices of measurement settings made by the experimenters are also determined given the initial conditions. Note that this is just a consequence of applying determinism to the entire universe.\\
2. Measurement dependence: The hidden variables $\lambda$ are statistically correlated with the measurement settings $M$ in general. In addition, it is usually assumed that this correlation is such that the Bell correlations are reproduced.\\

The first property, determinism, implies that the second property, measurement dependence, must be encoded in the past conditions. The past conditions must posit an appropriate correlation between $\lambda$ and the mechanism that determines the measurement settings so that $\lambda$ and the measurement settings are correlated. For example, the measurement settings might be determined by the wavelength of photons emitted by a distant quasar. Then, the photon emission process at the quasar, and thereby the wavelength of the photons emitted, has to be correlated with $\lambda$ for there to be a correlation between $\lambda$ and the measurement settings. This correlation is posited to be a consequence of the initial conditions of the universe, which determines both the wavelength of the photons and the $\lambda$'s. Some authors have taken the defining feature of superdeterminism to be a correlation between $\lambda$ and the factors that determine the measurement settings (without assuming determinism) \cite{woodspek}. \\

Both the properties are independent of each other. For example, pilot-wave theory \cite{bohm1, bohm2, solventini} is deterministic but not measurement dependent in general. Both the setting mechanism and the quantum system can be deterministically described by the pilot-wave theory. But if the setting mechanism and the quantum system are not entangled, the measurement settings and the hidden variables that determine the measurement outcomes will not be correlated. On the other hand, retrocausal hidden-variable models \cite{costacoffee, cramer, pricebook, sutherland, whartonmain, fpaper, seneffect, whartreview} are measurement dependent but not deterministic. These models posit that events are not fully determined by past conditions alone, but that a full determination requires the specification of future boundary conditions as well. The future conditions, in these models, encode the measurement settings. The hidden-variables distribution is then said to be, in some sense, causally affected by the measurement settings backwards in time. As these models are not deterministic (given the past conditions alone), they are not superdeterministic. As of yet, hidden-variables models which violate measurement-independence are either superdeterministic or retrocausal. In this paper, we exclusively focus on superdeterministic models.\\

It is useful to distinguish between two types of superdeterministic models, which we may call type I and type II. In a type I model, the correlation between $\lambda$ and the setting mechanism can be explained in terms of either past common causes or causal influences between them. In a type II model, the correlation is explained as a direct consequence of the initial conditions of the universe. In this case, events having no common past can also be correlated given appropriate initial conditions\footnote{In some type II models the initial conditions can be subject to important constraints. Such a model has, for example, been proposed by Palmer \cite{palmer09, palmer1, palmer2, palmerend}. See ref. \cite{sinp} for a detailed development and analysis of Palmer's proposal as a hidden-variables model.} (see Fig. 1). To illustrate the difference between the two types, consider the recent experiment \cite{cosmicbellII} where photons from $7.78$ billion years ago were used to choose the measurement settings for a Bell experiment. Consider the events corresponding to the photon emission and the experiment. The overlap in the past lightcones of these events comprised $\sim 4 \%$ of the total space-time volume of the past lightcone of the experiment. In a type I model of the experiment, the correlation between the hidden variables and the measurement settings originates exclusively from this tiny space-time volume of the common past. In a type II model however, the correlation arises from the initial conditions of the entire universe at the time of the big bang. Therefore, the experiment does not significantly constrain type II superdeterministic models.\\

\begin{figure}\label{A}
\includegraphics[width= 16cm]{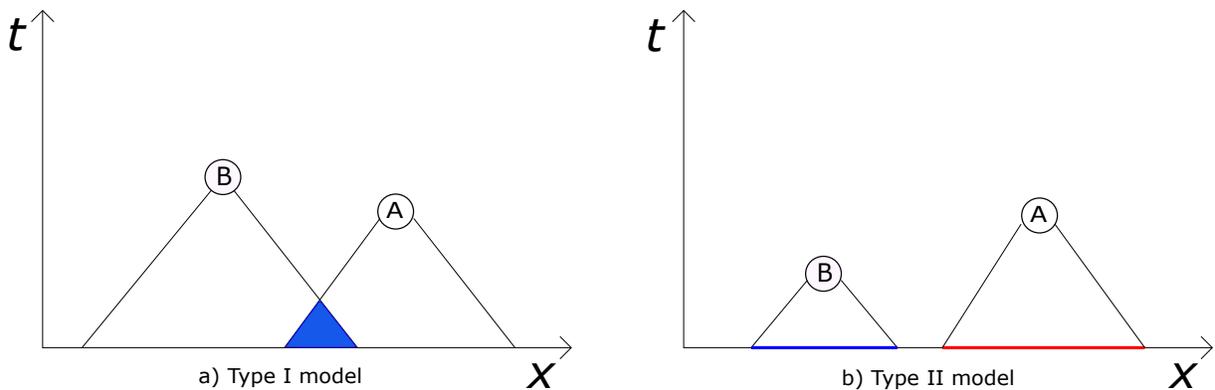}
\caption{The two classes of superdeterministic models. The time is mapped on the $y$-axis and the distance on the $x$-axis. The big bang singularity is mapped onto the $t=0$ surface. Part $a)$ of the figure depicts a type I model, where the correlation between two events A and B is either due to common causes in their common past (shaded in blue) or due to a causal connection from A to B (if B lies within the forward lightcone of A). Part $b)$ of the figure depicts a type II model, where two events A and B are correlated due to the initial conditions at $t=0$, without any need for common causes or causal connections between them. The information about the initial conditions has been divided into the past exclusive to A (shaded in red), and the past exclusive to B (shaded in blue). In a type II model, the information shaded in blue is correlated with the information shaded in red such that the events A and B are correlated.}
\end{figure}

Superdeterministic models circumvent Bell's theorem and reproduce the Bell correlations in a local manner. However, these models have been widely criticised in the literature \cite{dialect, pricebash, pironio}. These criticisms can be broadly divided into two arguments. The first argument, which is directed against determinism, is that they conflict with our apparent sensation of `free-will', since the experimenters' choices of measurement settings are completely determined from past conditions in these models. This criticism is based on the common misconception that human volition (or `free-will') can be explained by indeterminism. But in fact indeterminism arguably makes `free-will' even harder to explain, since the choices an experimenter makes would then have no cause at all. If an agent makes a choice for no reason, then the agent should be surprised by the choice - since nothing in the past, not even the agent's own thoughts and feelings, can explain it \cite{scruton}. Such misconceptions about `free-will' are also arguably undermined by recent advances in neuroscience, which appear to demonstrate that a human subject's choice is encoded in their brain activity up to 10 s before the choice enters the subject's conscious awareness \cite{neuro}. \\

It has sometimes been argued that the assumption of `free-will' is essential for the scientific method. For example, Zeilinger writes ``This is the assumption of `free-will'. It is a free decision what measurement one wants to perform...This fundamental assumption is essential to doing science. If this were not true, then, I suggest, it would make no sense at all to ask nature questions in an experiment, since then nature could determine what our questions are, and that could guide our questions such that we arrive at a false picture of nature'' \cite{zeilbook}. The argument clearly brings out a tension between the two basic assumptions of science: first, that nature is described by laws; second, that these laws can be experimentally tested. This tension arises because the experimenters are themselves part of nature, and thus described by the same laws. However, the solution cannot be to uncritically fall back upon `free-will', as that violates the first assumption. There needs to be an in-depth philosophical enquiry into the scientific implications of abandoning the indeterministic notion of `free-will'. Pending such an enquiry, it is premature to describe superdeterministic models as unscientific.\\

Recently, Hardy \cite{hardyjoke} has proposed testing a hybrid model where the universe is local and superdeterministic except for conscious human minds which are assumed to have `free-will'; that is, human choices introduce genuinely new information into the universe in this model.  This information then spreads out from the event location at a speed equal to or less than that of light. The model predicts that the Bell inequalities will be violated in all cases except when human beings are used to choose the measurement settings. Recently a group of researchers, called `The Big Bell Test Collaboration', used humans to choose the measurement settings for a Bell experiment \cite{faltunature}. The motivation for the experiment was to test such a hybrid model of the universe (although the assumptions of the model were less unambiguously stated than by Hardy). They found that Bell inequalities are violated even when humans choose the measurement settings. The hybrid model is therefore falsified by the experiment; but it is not clear which part of the model is the culprit: the assumption that humans have `free-will', or that the rest of the universe is local and superdeterministic.\\ 

The second argument is that superdeterministic models are `conspiratorial'. The charge of conspiracy against such models was made by Bell, who wrote \cite{bertlmann}
\begin{quote}``Now even if we have arranged that [the measurement settings] $a$ and $b$ are generated by apparently random radioactive devices, housed in separate boxes and thickly shielded, or by Swiss national lottery machines, or by elaborate computer programmes, or by apparently free willed experimental physicists, or by some combination of all of these, we cannot be sure that $a$ and $b$ are not significantly influenced by the same factors $\lambda$ that influence [the measurement results] $A$ and $B$. But this way of arranging quantum mechanical correlations would be even more mind boggling than one in which causal chains go faster than light. Apparently separate parts of the world would be deeply and conspiratorially entangled...''\footnote{In an earlier exchange, however, Bell appeared somewhat more open to this possibility \cite{bellisfree}: ``A theory may appear in which such conspiracies inevitably occur, and these conspiracies may then seem more digestible than the non-localities of other theories. When that theory is announced I will not refuse to listen, either on methodological or other grounds.''}
\end{quote}

The purpose of this article and $\mathcal{B}$ is to further develop the conspiracy argument by making it mathematically concrete. In this article we lay the groundwork by a study of the properties of superdeterministic models in nonequilibrium. In $\mathcal{B}$ we use our results from this study to quantitatively develop the notion of conspiracy. The present article is structured as follows. We first introduce nonequilibrium extensions of superdeterministic models in section \ref{omega}. An intuitive definition of conspiracy based on fine-tuning is immediately suggested. The section also points out that the formal no-signalling constraints fail to capture the physical meaning of signalling for these models. In section \ref{omega2}, we review the mechanism by which nonlocal deterministic models violate formal no-signalling \cite{genon} (in nonequilibrium). Using a similar approach, we develop a mechanism for superdeterministic models in section \ref{sd}. In section \ref{omega3}, we discuss another conspiratorial feature of superdeterminism that emerges in signalling. We conclude with a discussion in section \ref{omega4}.\\

\section{Nonequilibrium and signalling in superdeterministic models}\label{omega}
A scientific theory has two logically separate components: the initial or boundary conditions (which are contingent) and the laws (which are immutable). Delineating these two in a hidden-variables model naturally leads to the concept of nonequilibrium, as follows. Consider a hidden-variables model of a single quantum system (which may be composed of several quantum particles) as opposed to an ensemble of quantum systems. The model will attribute an initial $\lambda$ to the system. The initial $\lambda$ is a contingent feature of the model, as it is an initial condition. On the other hand, the mapping from $\lambda$ to the measurement outcomes (given the measurement settings) is a law-like feature of the model. Now consider a hidden-variables model of an ensemble of quantum systems. The model will attribute a hidden-variables distribution to the ensemble. As the initial $\lambda$ is contingent for each system, it follows that the initial hidden-variables distribution is contingent for the ensemble. A particular distribution (the `equilibrium' distribution) reproduces the quantum predictions, but the model inherently contains the possibility of other `nonequilibrium' distributions.\footnote{It has been argued by D{\"u}rr et al. \cite{92mainstream} that in pilot-wave theory the equilibrium distribution is typical (with respect to the equilibrium measure), and that this rules out nonequilibrium distributions. However, while the equilibrium distribution is indeed typical with respect to the equilibrium measure, it is also untypical with respect to a nonequilibrium measure \cite{teenv}. Thus, the argument is really circular.} \\

In some theories the initial values of the hidden variables may be subject to important constraints. This is commonplace even in classical physics. For example, an initial electromagnetic field must satisfy the constraints $\vec{\nabla}\cdot\vec{E} = \rho$ and $\vec{\nabla}\cdot \vec{B}=0$ (where $\rho$ is the charge density). The other two Maxwell equations contain time derivatives and determine the dynamical evolution from the initial conditions. The situation is similar in general relativity: only the space-space components of the Einstein field equations are truly dynamical, the other components define constraints on an initial spacelike slice. When considering theories with constraints on the state space, there are three points to note. First, such constraints are best viewed as part of the definition of the physical state space (that is, the space of allowed states) for an individual system. Second, for a given individual system the actual physical initial state (satisfying the constraints) is a contingency. Third, for an ensemble of systems, the initial probability distribution over the physical state space is likewise a contingency. Thus, considering theories with a constrained state space in no way affects the argument for the contingency of quantum equilibrium in hidden-variables theories generally.\\

The concept of quantum nonequilibrium was first proposed in the context of pilot-wave theory and used to study the issue of signal locality \cite{valentinI, valentinII}. However, the concept itself does not depend on the details of any particular theory or on the particular issue being addressed. Subsequent study of nonequilibrium for general nonlocal models showed that the usual explanation of signal locality involves fine tuning of the intial distribution \cite{genon}. Note that the same conclusion was arrived at later by other workers applying causal discovery algorithms to Bell correlations \cite{woodspek}. Lastly, nonequilibrium and signal locality have also been studied for retrocausal models \cite{fpaper}.\\

For a superdeterministic model, the novel addition is the contingent nature of the correlation between $\lambda$ and the measurement settings. As discussed in the Introduction, this correlation is a consequence of the correlation between $\lambda$ and the setting mechanism. The latter correlation is contingent (determined by the initial conditions) and therefore variable in principle. In general, $\lambda$ may be correlated with each setting mechanism in a different way. Suppose there are two setting mechanisms at each wing of a Bell experiment. Say there are two different computer algorithms that generate possible setting values for the experimenter to choose. Let us label the setting values generated by the two setting mechanisms at wing A (B) by the variables $\alpha_1$ and $\alpha_2$ ($\beta_1$ and $\beta_2$). The hidden-variables distribution will then be given by\footnote{The dependence of the hidden-variables distribution on the quantum state is implicit, and is suppressed hereafter for convenience.} $\rho(\lambda| \alpha_i, \beta_j)$, where $i, j \in \{1, 2\}$. Suppose the measurement setting at wing A (B) is $M_A$ ($M_B$). There will then be four logically independent distributions $\rho(\lambda| \alpha_i = M_A, \beta_j = M_B)$. In general, these will be different, giving rise to different measurement statistics (for the same settings $M_A$ and $M_B$). Thus, \textit{the choice of setting mechanism can affect the measurement statistics for a nonequilibrium extension of a superdeterministic model}. We intuitively expect that superdeterministic models have to be fine-tuned so that the measurement statistics do not depend on the choice of setting mechanism. In $\mathcal{B}$ we formulate this intuition quantitatively.\\

Let us now consider the signalling properties of these models. Suppose the setting mechanisms $i$ and $j$ are used. The model will predict the expectation values
\begin{align}
E[AB|M_A, M_B] &= \int d\lambda A(\lambda,M_A) B(\lambda, M_B) \rho(\lambda|\alpha_i= M_A, \beta_j = M_B) \label{ij}\\
E[A|M_A, M_B] &= \int d\lambda A(\lambda, M_A) \rho(\lambda|\alpha_i = M_A, \beta_j = M_B)  \label{i}\\
E[B|M_A, M_B] &= \int d\lambda B(\lambda, M_B) \rho(\lambda|\alpha_i =M_A, \beta_j = M_B) \label{j}
\end{align}
where $A(\lambda,M_A)$ and $B(\lambda, M_B)$ (called `indicator functions') determine the local measurement outcomes at wings A and B respectively. In general, the equations (\ref{ij}), (\ref{i}) and (\ref{j}) will not match quantum predictions if $\rho(\lambda| \alpha_i=M_A, \beta_j=M_B) \neq \rho_{eq}(\lambda| M_A, M_B)$. In that case, since the Bell correlations satisfy the formal no-signalling constraints, one intuitively expects that equations (\ref{i}) and (\ref{j}) will violate them in general.\\

It is useful to point out here that the notion of no-signalling was criticised by Bell \cite{cuisine} as resting on concepts ``which are desperately vague, or vaguely applicable". His criticism was based on the grounds that ``the assertion that `we cannot signal faster than light' immediately
provokes the question: Who do we think \textit{we} are?''. This anthropocentric criticism is brought to a head in the context of superdeterminism, where the experimenters' choices of measurement settings are explicitly considered as variables internal to the model (as the outputs of the setting mechanisms). For a superdeterministic model, the violation of formal no-signalling is not equivalent to actual signalling. To make the discussion precise, we define an actual signal to be present (say, from $M_B \to A$) in a Bell experiment only if:\\
1. The formal no-signalling constraints are violated. That is, $p(A| M_A, M_B) \neq p(A| M_A, M_B')$, where $A$ is the local measurement outcome at the first wing. \\
2. The distant measurement setting $M_B$ is a cause of the local outcome $A$. For a deterministic model (as considered here), we define that $A$ causally depends on $M_B$ if only if $A$ functionally depends on $M_B$. \\

If only formal no-signalling is violated, then we refer to this violation as an apparent signal (in contrast to an actual signal). The causality condition is needed to ensure that we do not mistake a statistical correlation between $A$ and $M_B$ for actual signalling (see also section \ref{omega3} for a discussion of causality for indeterministic models). The importance of this condition is brought out clearly by superdeterministic models in nonequilibrium. Say the measurement setting at wing B is changed from $M_B \rightarrow M_B'$. The hidden-variables distribution will change from $\rho(\lambda|\alpha_i = M_A, \beta_j = M_B) \rightarrow \rho(\lambda|\alpha_i = M_A, \beta_j = M_B')$ -- not due to a causal effect of the change $M_B \rightarrow M_B'$, but because the distribution is statistically correlated (due to past initial conditions) with the setting mechanism output $\beta_j$. That is, the hidden-variables distribution does not functionally depend on the measurement settings; it is only correlated with them. The local indicator function $A(\lambda, M_A)$ is also functionally independent of $M_B$. This is true in both equilibrium and in nonequilibrium. The violation of formal no-signalling in nonequilibrium (analogous to that of Bell inequalities in equilibrium) arises as a peculiarity of the statistical correlation between the setting mechanism and $\lambda$. Thus, the violation of formal no-signalling constraints in superdeterminism is an apparent signal. Since the violation of formal no-signalling constraints is a necessary but not sufficient condition for actual signalling, \textit{the so-called no-signalling constraints may be more appropriately called `marginal-independence' constraints}. \\

In section \ref{omega3}, we discuss this further in the context of superdeterministic conspiracy. Currently, we pivot our attention to the hidden-variables mechanism by which marginal-independence is violated. In the next section we review this mechanism for nonlocal deterministic models, as discussed in ref. \cite{genon}. 
\begin{figure}
\includegraphics[scale=0.7]{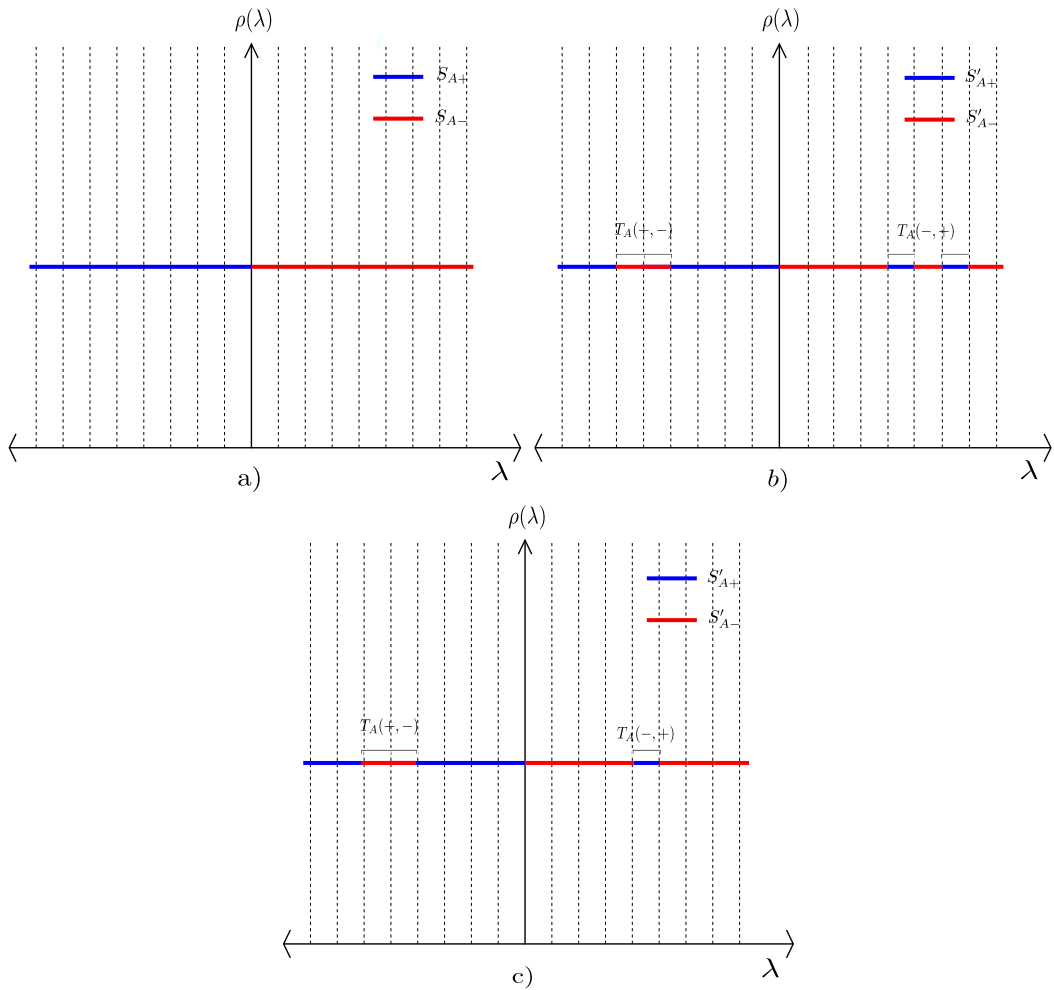}
\caption{Schematic illustration of signalling in nonlocal models (following the methodology of ref. \cite{genon}). Part $a)$ of the figure represents the initial case where the measurement settings at the two wings are $M_A$ and $M_B$. The hidden-variables distribution has been assumed to be uniform for simplicity. Upon changing $M_B$ $\rightarrow M_B'$ some of the $\lambda$'s which initially gave $A=+1$($-1$) `flip their outcomes' to subsequently give $A= -1$($+1$) due to nonlocality. If the distribution satisfies marginal-independence, for example the equilibrium distribution $\rho_{eq}(\lambda)$, then the set of $\lambda$'s that flip their outcomes from $+1\rightarrow  -1$, represented by the `transition set' $T_A(+,-)$, has the same measure as the set of $\lambda$'s that flip their outcomes from $-1\rightarrow +1$, represented by $T_A(-,+)$. This case is illustrated in part $b)$ of the figure, where the measures of  $T_A(+,-)$ and $T_A(-,+)$ are equal. For a nonequilibrium distribution however, the measure of the two transition sets are not equal in general, which leads to signalling. This is illustrated in part $c)$ of the figure. Here $S_{A+(-)}=\{\lambda| A(\lambda, M_A, M_B) = +(-)1\}$ and $S'_{A+(-)}=\{\lambda| A(\lambda, M_A, M_B') = +(-)1\}$.}
\end{figure}

\section{Marginal-independence in nonlocal deterministic models}\label{omega2}
Consider the standard Bell scenario \cite{bell1975}, where two spin-1/2 particles are prepared in the spin-singlet state and subsequently subjected to the local measurements $M_A$ (corresponding to the operator $ \hat{\sigma}_{\hat{a}}\otimes \hat{I}$) and $M_B$ (corresponding to the operator $\hat{I} \otimes \hat{\sigma}_{\hat{b}}$) in a spacelike separated manner. A nonlocal deterministic (but not superdeterministic) model, which specifies the measurement outcomes by the nonlocal indicator functions $A(\lambda, M_A, M_B)$ and $B(\lambda, M_A, M_B)$, reproduces the Bell correlations for an `equilibrium' distribution of hidden variables $\rho_{eq}(\lambda)$. Consider nonequilibrium distributions $\rho(\lambda) \neq \rho_{eq}(\lambda)$ for the model (with the same indicator functions). The joint outcome expectation value can be expanded as

\begin{align}
E[AB|M_A, M_B] = \int d\lambda A(\lambda, M_A, M_B) B(\lambda, M_A, M_B) \rho(\lambda)
\end{align}
while the marginal outcome distributions can be expanded as
\begin{align}
E[A|M_A, M_B] &=  \int d\lambda A(\lambda, M_A, M_B) \rho(\lambda) \label{10}\\
E[B|M_A, M_B] &= \int d\lambda B(\lambda, M_A, M_B) \rho(\lambda) \label{10'}
\end{align}
Equations (\ref{10}) and (\ref{10'}) depend on the measurement settings at both wings. Thus, one intuitively expects that a change in one of the measurement settings will in general change both the marginal outcome distributions, thus violating marginal-independence\footnote{In section \ref{omega3}, we discuss that the violation of marginal-independence implies an actual signal for nonlocal models. Therefore, we currently use the term signal when these constraints are violated.}. It can however be that, for some special hidden-variables distributions, changing a measurement setting affects the marginal outcome distribution only at that same wing, as is true, for example, for the equilibrium distribution $\rho_{eq}(\lambda)$. \\

This intuition can be formalised using the methodology of ref. \cite{genon}. Consider changing the measurement setting $M_B$ (corresponding to $\hat{\sigma}_{\hat{b}}$) $\rightarrow M_B'$ (corresponding to $\hat{\sigma}_{\hat{b}'}$). Consider the set $S=\{\lambda| \rho(\lambda) >0\}$. We may define the subsets $S_{A+}=\{\lambda| \lambda \in S \wedge A(\lambda, M_A, M_B) = +1\}$, $S_{A-}=\{\lambda| \lambda \in S \wedge A(\lambda, M_A, M_B) = -1\}$ and  $S'_{A+}=\{\lambda| \lambda \in S \wedge A(\lambda, M_A, M_B') = +1\}$, $S'_{A-}=\{\lambda| \lambda \in S \wedge A(\lambda, M_A, M_B') = -1\}$. Clearly $S_{A+} \cap S_{A-} = S'_{A+} \cap S'_{A-} = \emptyset$, and the set $S$ can be partitioned as $S = S_{A+} \cup S_{A-} =S'_{A+} \cup S'_{A-}$. We may also define the `transition sets' $T_A(+,-) = S_{A+}\cap S'_{A-}\big ($the set of $\lambda$'s that transition from giving $A=+1$ to $A=-1$ upon changing $M_B$ $\rightarrow M_B'\big )$ and $T_A(-,+) = S_{A-}\cap S'_{A+}$ $\big ($the set of $\lambda$'s that transition from giving $A=-1$ to $A=+1$ upon changing $M_B$ $\rightarrow M_B'\big )$. The property that an arbitrary distribution $ \rho(\lambda)$ must possess to be marginal-independent can then be expressed as simply the equality of the measures of the two transition sets (see Fig. 2), that is
\begin{align}
\int_{T_A(+,-)} d\lambda \rho(\lambda) = \int_{T_A(-,+)} d\lambda \rho(\lambda)
\end{align}
This equality means that the fraction of the ensemble making a transition from $+1$ $\rightarrow -1$ is equal to the fraction of the ensemble making the reverse transition (a form of `detailed balancing'), so that the marginal distribution at A is unchanged (under a change of measurement setting at B). On the other hand, if this equality is violated for the given settings $M_A$, $M_B$, $M_B'$ -- as will be the case for a general nonequilibrium distribution -- then for those settings the marginal at A will change and there will be a nonlocal signal from B to A \cite{genon}. The change in the marginal at A, in this case, will be
\begin{align}
&\int_{T_A(-,+)} d\lambda \rho(\lambda) - \int_{T_A(+,-)} d\lambda \rho(\lambda) \\
= &\int_{S_{A-}\cap S'_{A+}} d\lambda \rho(\lambda) - \int_{S_{A+}\cap S'_{A-}} d\lambda \rho(\lambda)\\
= &\bigg (\int_{S_{A-}\cap S'_{A+}} d\lambda \rho(\lambda) + \int_{S_{A+}\cap S'_{A+}} d\lambda \rho(\lambda) \bigg ) - \bigg ( \int_{S_{A+}\cap S'_{A-}} d\lambda \rho(\lambda) + \int_{S_{A+}\cap S'_{A+}} d\lambda \rho(\lambda) \bigg )\\
= &\int_{S'_{A+}} d\lambda \rho(\lambda) - \int_{S_{A+}} d\lambda \rho(\lambda) \label{arb}
\end{align}
where, in the last line, we have used the relations $(S_{A-}\cap S'_{A+}) \cup (S_{A+}\cap S'_{A+}) = S'_{A+}$ and $(S_{A+}\cap S'_{A-}) \cup (S_{A+}\cap S'_{A+}) = S_{A+}$. In the next section, we use a similar methodology to analyse marginal-independence in superdeterministic models. 

\section{Marginal-independence in superdeterministic models}\label{sd}
Consider a local superdeterministic model of the Bell scenario, where the measurement outcomes are specified by the local indicator functions $A(\lambda, M_A)$ and $B(\lambda, M_B)$. Here $M_A$ ($M_B$) corresponds to the operator $\hat{\sigma}_{\hat{a}}\otimes \hat{I}$ ($\hat{I}\otimes\hat{\sigma}_{\hat{b}}$). The model reproduces the Bell correlations for an equilibrium distribution of hidden variables $\rho_{eq}(\lambda|M_A, M_B)$. Consider a nonequilibrium distribution $\rho(\lambda|\alpha = M_A, \beta = M_B) \neq \rho_{eq}(\lambda|M_A, M_B)$, where $\alpha$ ($\beta$) labels the output of the setting mechanism that determines the measurement setting at wing A (B).\\

Let us examine the details at the hidden-variables level when a measurement setting is changed, say $M_B$ $\rightarrow M_B'$.
\begin{figure}
\includegraphics[scale=0.7]{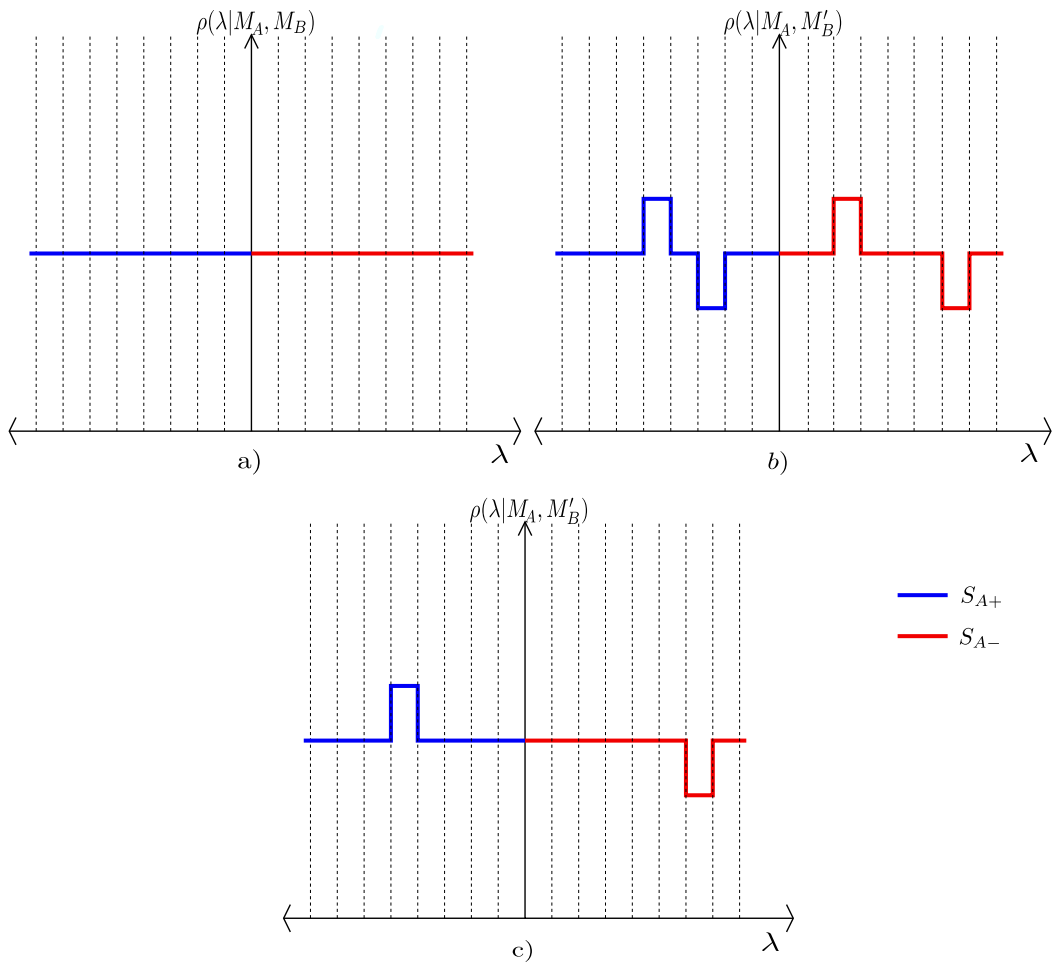}
\caption{Schematic illustration of apparent-signalling in superdeterministic models. Part $a)$ of the figure shows the hidden-variables distribution, assumed to be uniform for simplicity, when the measurement settings are $M_A$ and $M_B$ at wings A and B respectively. The blue portion of the graph represents $\lambda$'s belonging to the set $S_{A+}=\{\lambda|\lambda \in S \wedge A(\lambda, M_A) = +1\}$, and the red portion represents $\lambda$'s belonging to the set $S_{A-}=\{\lambda|\lambda \in S \wedge A(\lambda, M_A) = -1\}$.  If the hidden-variables distribution is marginal-independent, then a change $M_B$ $\rightarrow M_{B'}$ will correspond to the measures of the sets $S_{A+}$ and $S_{A-}$ remaining constant. This is illustrated in part $b)$, where the measures of the sets $S_{A+}$ and $S_{A-}$ remain unchanged, although the distribution has changed from $\rho(\lambda|\alpha=M_A, \beta=M_B) \rightarrow \rho(\lambda|\alpha=M_A, \beta=M_B')$. In general, for a nonequilibrium distribution, a change in $M_B$ will correspond to a change in the measures of $S_{A+}$ and $S_{A-}$. This is illustrated in part $c)$ of the figure.}
\end{figure}
We define the sets $S_{M_B} =\{\lambda|\rho(\lambda|\alpha=M_A, \beta=M_B) > 0\}$, $S_{M_B'} =\{\lambda|\rho(\lambda|\alpha=M_A, \beta=M_B') >0\}$, $S = S_{M_B} \cup S_{M_B'}$ and partition the set $S$ as
\begin{align}
S = S_{A+}\cup S_{A-}; \textbf{ } \textbf{ }  S_{A+}\cap S_{A-} =\emptyset \label{y}
\end{align}
where $S_{A+} =\{\lambda|\lambda \in S \wedge A(\lambda, M_A) = +1\}$ and $S_{A-} =\{\lambda|\lambda \in S \wedge A(\lambda, M_A) = -1\}$. The measure of the set $S_{A+}$ with respect to the distribution $\rho(\lambda|\alpha=M_A, \beta=M_B)$ $\big ( \rho(\lambda|\alpha=M_A, \beta=M_B')\big )$ determines the marginal probability that $A=+1$ when the measurement setting at wing B is $M_B$ ($M_B'$). For marginal-independence, we must have
\begin{align}
\int_{S_{A+}} d\lambda \rho(\lambda|\alpha=M_A, \beta=M_B) = \int_{S_{A+}} d\lambda \rho(\lambda|\alpha=M_A, \beta=M_B') \label{x}
\end{align}
Since the measure of the set $S$ is 1 with respect to both the distributions, we can use equations (\ref{y}) and (\ref{x}) to get
\begin{align}
\int_{S_{A-}} d\lambda \rho(\lambda|\alpha=M_A, \beta= M_B) = \int_{S_{A-}} d\lambda \rho(\lambda|\alpha=M_A, \beta=M_B') \label{z}
\end{align}
Equations (\ref{x}) and (\ref{z}) can be restated together as
\begin{align}
\int_{S_{A+(-)}} d\lambda \rho(\lambda|\alpha=M_A, \beta=M_B) = \int_{S_{A+(-)}} d\lambda \rho(\lambda|\alpha=M_A, \beta=M_B') 
\end{align}
Thus, for marginal-independence, the measures of the sets $S_{A+}$ and $S_{A-}$ must remain constant when $M_B$ is changed to $M_B'$. For nonequilibrium distributions, the measures will change in general (see Fig. 3). This can be intuitively thought of in the following manner. Changing $M_B$ $\rightarrow M_B'$ corresponds to a `reshuffle' of the distribution $\rho(\lambda|\alpha=M_A, \beta=M_B)$ in the set $S$ to $\rho(\lambda|\alpha=M_A, \beta=M_B')$. Reshuffling implies that if the distribution is increased at some region of $\lambda$, then it has to be equally decreased at another region to keep the total measure constant. In the case of the equilibrium distribution, the sets $S_{A+}$ and $S_{A-}$ are reshuffled separately. This keeps the measures of $S_{A+}$ and $S_{A-}$ constant, although the distribution changes inside both the sets. In general, the distribution is reshuffled \textit{across the entire set $S$}. The measures of $S_{A+}$ and $S_{A-}$ therefore change in general, violating marginal-independence. The change in the marginal at A, in this case, will be given by
\begin{align}
\int_{S_{A+}} d\lambda \rho(\lambda|\alpha=M_A, \beta=M_B') - \int_{S_{A+}} d\lambda \rho(\lambda|\alpha=M_A, \beta=M_B) \label{arby}
\end{align}
Comparing this with equation (\ref{arb}), it becomes evident that this mechanism is completely different from that discussed for nonlocal deterministic models in the previous section. In the case of nonlocality, marginal-independence is violated due to inequality of the measures of the transition sets $T_A(-,+)$ and $T_A(+,-)$. There are no transition sets in the superdeterministic case. The mapping from $\lambda$ to the measurement outcomes is not affected by the change of measurement settings -- the distribution of $\lambda$ itself changes. The change in marginals (in nonequilibrium) occurs as the measures of $S_{A+}$ and $S_{A-}$ change when $M_B$ is changed to $M'_{B}$. 

\section{The conspiratorial character of superdeterministic signalling}\label{omega3}
In section \ref{omega}, we discussed that superdeterministic signalling is apparent, that is, although marginal-independence (formal no-signalling) is violated there is no causal relationship between the local outcomes and the distant measurement settings. In this section, we argue, first, that apparent-signalling is a unique feature of superdeterministic models, and second, that it is indicative of the conspiratorial character\footnote{Note that we use the word ``conspiracy'' to address the criticism by Bell about superdeterministic models (see the Introduction). Conspiracy can also be used, in a different context, to indicate that hidden-variables models need fine tuning to reproduce marginal-independence \cite{genon, woodspek}. The usages of the same word are different. In this article and in $\mathcal{B}$, we are concerned only with Bell's notion of superdeterministic conspiracy.} of these models. \\

We argue for the uniqueness by considering violation of formal no-signalling vis-a-vis actual signalling for other hidden-variables models. Let us first consider deterministic nonlocal models. The outcome at the first wing $A$ is a function of the measurement setting $M_B$ at the second wing. Therefore, there is a causal link $M_B \to A$. For nonlocal models, then, the violation of marginal-independence in nonequilibrium constitutes an actual signal. Next, consider retrocausal models. For these models, the definition of causality given in section \ref{omega} is insufficient as retrocausal models are indeterministic (given the past conditions alone). For indeterministic models, we make use of the causal modelling framework (interventionist causation) \cite{jearl}, where a variable $x$ is considered to be a cause of $y$ if only if an external intervention on $x$ changes $y$ but not vice versa. In the present context, this implies that $M_B$ has to be an ancestor of $A$ for $M_B$ to causally affect $A$.  In retrocausal models, $M_B$ is an ancestor of $A$ as there is a causal link $M_B \to \lambda \to A$. That is, $\lambda$ is causally affected by $M_B$ (backwards in time), then $\lambda$ affects $A$ (forwards in time). Therefore, for retrocausal models also, the violation of marginal-independence in nonequilibrium (see, for example, ref. \cite{fpaper}) constitutes an actual signal. Therefore, apparent-signalling is unique to superdeterministic models. This is illustrated in Fig. 4.\\

This apparent-signalling brings out another conspiratorial feature of superdeterminism. The experimenter at wing A (B) of the Bell scenario can influence the marginal outcome distribution at wing B (A), because marginal-independence is violated. Practically, this constitutes a signalling procedure, but the physical interpretation is that the local marginals and the distant settings are only statistically correlated, due to past initial conditions. This leads to a peculiar scenario where the entire sequence of messages exchanged between the two experimenters is explained as a statistical coincidence. Consider a superdeterministic model of spin-singlet pairs with a nonequilibrium hidden-variables distribution, for which the measurement statistics violate marginal-independence. The two experimenters may exploit this to construct an `instantaneous telephone' to communicate with each other. Although there is no causal link between the wings in a superdeterministic model, the fact remains that marginal-independence is violated, which the experimenters may exploit for communication. Say the experimenter at the first wing speaks a message A into the telephone. The experimenter at the second wing hears the message A at their end, but this is, according to superdeterminism, only a coincidence. In principle, the experimenter could have heard a message A', but A'=A is \textit{arranged} by the past initial conditions. Say the second experimenter replies back with a message B. The first experimenter then hears the correct message B, but this is again purely coincidental. The two experimenters can continue to have an arbitrarily long conversation using the telephone, without any information being exchanged between the wings. \\

A `series of coincidences mimicking an actual conversation' appears conspiratorial intuitively. In $\mathcal{B}$ we show that such series of coincidences are endemic in superdeterminism, and we quantify the conspiracy involved. \\

\begin{figure}
\includegraphics[scale=0.5]{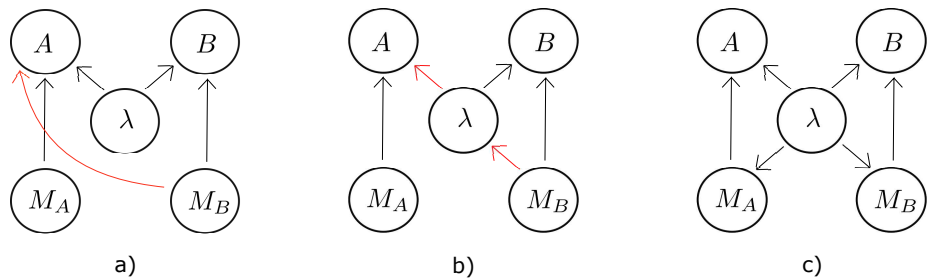}
\caption{Schematic illustration of the unique conspiratorial character of superdeterministic signalling. Parts $a)$, $b)$ and $c)$ illustrate nonlocal, retrocausal and superdeterministic models of the Bell correlations. For nonlocal models, the distant measurement setting $M_B$ directly affects the local outcome $A$. For retrocausal models, the measurement setting $M_B$ indirectly affects $A$ via the route $M_B \to \lambda \to A$. The causal influence from $M_B \to \lambda$ is backwards in time, and that from $\lambda \to A$ is forwards in time. There is, however, no causal influence from $M_B \to A$ for a superdeterministic model. Therefore, although the violation of marginal-independence (formal no-signalling) indicates an actual signal in nonlocal and retrocausal models, it indicates only a statistical correlation between local outcomes and distant settings in superdeterminism. Note that, in part $c)$, $\lambda$ either causally influences the measurement settings or there is a common cause $\mu$ between them. Both possibilities are represented by the same figure as $\mu$ can be absorbed into the definition of $\lambda$.}
\end{figure}

\section{Conclusion}\label{omega4}
In an effort to isolate the key conspiratorial features of superdeterministic models, we have introduced nonequilibrium extensions of such models and explored their various properties. Our exercise has provided us with two leads. First, in section \ref{omega} we showed that different setting mechanisms in general lead to different measurement statistics in nonequilibrium. This implies that the model must be fine-tuned so that such effects do not appear in practice. Second, we showed in section \ref{omega3} that, in nonequilibrium, correlations can conspire to resemble an actual signal. In $\mathcal{B}$, we use these results to develop two separate ways to quantify the conspiratorial character of superdeterminism. \\

We also found that the formal no-signalling constraints fail to capture the intuitive idea of signalling for superdeterministic models. The violation of these constraints implies only a statistical correlation (due to initial conditions) between the local marginals and the distant settings. Therefore, we have suggested that the so-called no-signalling constraints may be more appropriately called marginal-independence constraints, and concluded that there is only an apparent-signal in superdeterministic models when these constraints are violated. The issue of signalling in superdeterministic hidden-variables models was also briefly raised in refs. \cite{hall16, scani}, where it was argued that the concept of signalling is not physically meaningful for these models (even if marginal-independence is violated). The argument is that the experimenters are not able to choose the measurement settings `at will' in such models. However, such reasoning would imply that there is no meaningful signalling in classical electrodynamics, which would run counter to our understanding of classical physics.\footnote{See also the Introduction for a discussion of the relationship between `will' and fundamental randomness.} We have argued that there are no actual signals in superdeterministic models for a different reason: although marginal-independence is violated in nonequilibrium, the local outcomes are not causally affected by the distant measurement settings. \\

Lastly, we have discussed the hidden-variables mechanism by which superdeterministic models violate marginal-independence, and we have distinguished this mechanism from that of nonlocal models. Thus, although hidden-variables models in general violate marginal-independence in nonequilibrium, the physical interpretation of this violation is model-dependent. \\

\acknowledgements
We are grateful to Tim Palmer and to the anonymous referees for helpful comments.

\bibliographystyle{bhak}
\bibliography{bib}

\end{document}